\documentclass[preprint,showpacs,superscriptaddress,nofootinbib]{revtex4}
  \usepackage{graphicx}
  \usepackage{rotating}
  \usepackage{multirow}
  \begin{document}
  \title{Study of the ${\psi}(1S,2S)$ and ${\eta}_{c}(1S,2S)$ weak decays into $DM$}
  \author{Junfeng Sun}
  \affiliation{Institute of Particle and Nuclear Physics,
              Henan Normal University, Xinxiang 453007, China}
  \author{Yueling Yang}
  \thanks{corresponding author}
  \email{yangyueling@htu.edu.cn}
  \affiliation{Institute of Particle and Nuclear Physics,
              Henan Normal University, Xinxiang 453007, China}
  \author{Jinshu Huang}
  \affiliation{College of Physics and Electronic Engineering,
              Nanyang Normal University, Nanyang 473061, China}
  \author{Lili Chen}
  \affiliation{Institute of Particle and Nuclear Physics,
              Henan Normal University, Xinxiang 453007, China}
  \author{Qin Chang}
  \affiliation{Institute of Particle and Nuclear Physics,
              Henan Normal University, Xinxiang 453007, China}

  \begin{abstract}
  Inspired by the recent measurements on the $J/{\psi}$ ${\to}$
  $D_{s}{\rho}$, $D_{u}K^{\ast}$ weak decays at BESIII and the
  potential prospects of charmonium at high-luminosity heavy-flavor
  experiments, we study ${\psi}(1S,2S)$ and ${\eta}_{c}(1S,2S)$ weak
  decays into final states including one charmed meson plus one
  light meson, considering QCD corrections to hadronic matrix
  elements with QCD factorization approach.
  It is found that the Cabibbo-favored ${\psi}(1S,2S)$
  ${\to}$ $D_{s}^{-}{\rho}^{+}$, $D_{s}^{-}{\pi}^{+}$,
  $\overline{D}_{u}^{0}\overline{K}^{{\ast}0}$ decays have
  large branching ratios ${\gtrsim}$ $10^{-10}$,
  which might be accessible at future experiments.
  \end{abstract}
  \maketitle

  \section{Introduction}
  \label{sec01}
  More than forty years after the discovery of the $J/{\psi}$
  meson, the properties of charmonium (bound state of $c\bar{c}$)
  continue to be the subject of intensive theoretical and
  experimental study.
  It is believed that charmonium, resembling bottomonium
  (bound state of $b\bar{b}$),
  plays the same role in exploring hadronic dynamics as
  positronium and/or the hydrogen atom plays in understanding the
  atomic physics. Charmonium and bottomonium are good objects
  to test the basic ideas of QCD \cite{phys.rept.1978}.
  There is a renewed interest in charmonium due to the
  plentiful dedicated investigation from BES, CLEO-c, LHCb
  and the studies via decays of the $B$ mesons at $B$ factories.

  The ${\psi}(1S,2S)$ and ${\eta}_{c}(1S,2S)$ mesons are
  $S$-wave charmonium states below open-charm kinematic threshold,
  and have the well-established quantum numbers of $I^{G}J^{PC}$
  $=$ $0^{+}1^{--}$ and $0^{+}0^{-+}$, respectively.
  They decay mainly through the strong and electromagnetic
  interactions.
  Because the $G$-parity conserving hadronic decays ${\psi}(2S)$
  ${\to}$ ${\pi}{\pi}J/{\psi}$, ${\eta}J/{\psi}$ and ${\eta}_{c}(2S)$
  ${\to}$ ${\pi}{\pi}{\eta}_{c}(1S)$ are suppressed by the compact
  phase space of final states, and because the decays into light
  hadrons are suppressed by the phenomenological Okubo-Zweig-Iizuka
  (OZI) rules \cite{o,z,i}, the total widths of ${\psi}(1S,2S)$
  and ${\eta}_{c}(1S,2S)$ are narrow (see Table \ref{tab:cc}),
  which might render the charmonium weak decay as
  a necessary supplement.
  Here, we will concentrate on the ${\psi}(1S,2S)$
  and ${\eta}_{c}(1S,2S)$ weak decays into $DM$ final states,
  where $M$ denotes the low-lying $SU(3)$ pseudoscalar
  and vector meson nonet.
  Our motivation is listed as follows.

   \begin{table}[h]
   \caption{The properties of the ${\psi}(1S,2S)$ and ${\eta}_{c}(1S,2S)$ mesons
   \cite{pdg}.}
   \label{tab:cc}
   \begin{ruledtabular}
   \begin{tabular}{cccc}
   meson & $I^{G}J^{PC}$ & mass (MeV) & width \\ \hline
   ${\psi}(1S)$ & $0^{+}1^{--}$ & $3096.916{\pm}0.011$ & $92.9{\pm}2.8$ keV \\
   ${\psi}(2S)$ & $0^{+}1^{--}$ & $3686.109^{+0.012}_{-0.014}$ & $299{\pm}8$ keV \\
   ${\eta}_{c}(1S)$ & $0^{+}0^{-+}$ & $2983.6{\pm}0.7$ & $32.2{\pm}0.9$ MeV \\
   ${\eta}_{c}(2S)$ & $0^{+}0^{-+}$ & $3639.4{\pm}1.3$ & $11.3^{+3.2}_{-2.9}$ MeV
   \end{tabular}
   \end{ruledtabular}
   \end{table}

  From the experimental point of view:
  (1)
  some $10^{9}$ ${\psi}(1S,2S)$ data samples have been
  collected by BESIII since 2009 \cite{bes3}.
  It is inspiringly expected to have about $10$ billion
  $J/{\psi}$ and 3 billion ${\psi}(2S)$ events at BESIII
  experiment per year of data taking with the designed
  luminosity \cite{cpc36};
  over $10^{10}$ $J/{\psi}$ at LHCb \cite{1509.00771},
  ATLAS \cite{npb850.387} and CMS \cite{prl114.191802}
  per ${\rm fb}^{-1}$ data in $pp$ collisions.
  A large amount of data sample offers a realistic
  possibility to explore experimentally the charmonium
  weak decays. Correspondingly, theoretical study is very
  necessary to provide a ready reference.
  (2)
  Identification of the single $D$ meson would provide an
  unambiguous signature of the charmonium weak decay into
  $DM$ states.
  With the improvements of experimental instrumentation and
  particle identification techniques, accurate measurements
  on the nonleptonic charmonium weak decay might be feasible.
  Recently, a search for the $J/{\psi}$ ${\to}$ $D_{s}{\rho}$,
  $D_{u}K^{\ast}$ decays has been performed at BESIII, although
  signals are unseen for the moment \cite{prd89.071101}.
  Of course, the branching ratios for the inclusive charmonium weak
  decay is tiny within the standard model, about
  $2/({\tau}_{D}{\Gamma}_{\psi})$ ${\sim}$ $10^{-8}$ and
  $2/({\tau}_{D}{\Gamma}_{{\eta}_{c}})$ ${\sim}$ $10^{-10}$,
  where $D$ denotes the neutral charmed meson \cite{zpc62},
  ${\Gamma}_{\psi}$ and ${\Gamma}_{{\eta}_{c}}$ stand for
  the total widths of the ${\psi}(1S,2S)$ and
  ${\eta}_{c}(1S,2S)$ resonances, respectively.
  Observation of an abnormally large production rate
  of single charmed mesons in the final state would be
  a hint of new physics beyond the standard model \cite{zpc62}.

  From the theoretical point of view:
  (1)
  The charm quark weak decay is more favorable than the bottom quark
  weak decay, because the Cabibbo-Kobayashi-Maskawa (CKM) matrix elements
  obey ${\vert}V_{cb}{\vert}$ ${\ll}$ ${\vert}V_{cs}{\vert}$ \cite{pdg}.
  Penguin and annihilation contributions to nonleptonic charm
  quark weak decay, being proportional to the CKM factor
  ${\vert}V_{cb}V_{ub}{\vert}$ ${\sim}$ ${\cal O}({\lambda}^{5})$
  with the Wolfenstein parameter ${\lambda}$ ${\simeq}$ $0.22$ \cite{pdg},
  are highly suppressed, and hence negligible relative to
  tree contributions.
  Both $c$ and $\bar{c}$ quarks in charmonium can decay individually,
  which provides a good place to investigate the dynamical mechanism
  of heavy flavor weak decay and crosscheck model parameters obtained
  from the charmed hadron weak decays.
  (2)
  There are few works devoted to nonleptonic $J/{\psi}$
  weak decays in the past, such as
  Ref. \cite{prd78} with the covariant light-cone quark model,
  Ref. \cite{epjc55} with QCD sum rules,
  and Refs. \cite{plb252,ijma14,adv2013}
  with the Wirbel-Stech-Bauer (WSB) model \cite{bsw}.
  Moreover, previous works of Refs. \cite{epjc55,prd78,plb252,ijma14,adv2013}
  concern mainly the weak transition form factors between
  the $J/{\psi}$ and charmed mesons.
  Fewer papers have been devoted to nonleptonic ${\psi}(2S)$
  and ${\eta}_{c}(1S,2S)$ weak decays until now even though
  a rough estimate of branching ratios is unavailable.
  In this paper, we will estimate the branching ratios
  for nonleptonic two-body charmonium weak decay,
  taking the nonfactorizable contributions to
  hadronic matrix elements into account with the
  attractive QCD factorization (QCDF) approach \cite{qcdf1}.

  This paper is organized as follows.
  In section \ref{sec02}, we will present the theoretical framework
  and the amplitudes for the ${\psi}(1S,2S)$, ${\eta}_{c}(1S,2S)$
  ${\to}$ $DM$ decays.
  Section \ref{sec03} is devoted to numerical results and discussion.
  Finally, section \ref{sec04} is our summation.

  \section{theoretical framework}
  \label{sec02}
  \subsection{The effective Hamiltonian}
  \label{sec0201}
  Phenomenologically,
  the effective Hamiltonian responsible for charmonium
  weak decay into $DM$ final states can be written
  as \cite{9512380}:
   \begin{equation}
  {\cal H}_{\rm eff}\ =\ \frac{G_{F}}{\sqrt{2}}\,
   \sum\limits_{q_{1},q_{2}}\, V_{cq_{1}}^{\ast} V_{uq_{2}}\,
   \Big\{ C_{1}({\mu})\,Q_{1}({\mu})
         +C_{2}({\mu})\,Q_{2}({\mu}) \Big\}
   + {\rm H.c.}
   \label{hamilton},
   \end{equation}
  where $G_{F}$ $=$ $1.166{\times}10^{-5}\,{\rm GeV}^{-2}$ \cite{pdg}
  is the Fermi coupling constant;
  $V_{cq_{1}}^{\ast}V_{uq_{2}}$ is the CKM factor with $q_{1,2}$ $=$ $d$, $s$;
  The Wilson coefficients $C_{1,2}(\mu)$, which are independent
  of one particular process, summarize the physical contributions
  above the scale of ${\mu}$.
  The expressions of the local tree four-quark operators are
    \begin{eqnarray}
    Q_{1} &=&
  [ \bar{q}_{1,{\alpha}}{\gamma}_{\mu}(1-{\gamma}_{5})c_{\alpha} ]
  [ \bar{u}_{\beta}{\gamma}^{\mu}(1-{\gamma}_{5})q_{2,{\beta}} ]
    \label{q1}, \\
    Q_{2} &=&
  [ \bar{q}_{1,{\alpha}}{\gamma}_{\mu}(1-{\gamma}_{5})c_{\beta} ]
  [ \bar{u}_{\beta}{\gamma}^{\mu}(1-{\gamma}_{5})q_{2,{\alpha}} ]
    \label{q2},
    \end{eqnarray}
  where ${\alpha}$ and ${\beta}$ are color indices.

  It is well known that
  the Wilson coefficients $C_{i}$ could be systematically calculated
  with perturbation theory and have properly been evaluated to
  the next-to-leading order (NLO).
  Their values at the scale of ${\mu}$ ${\sim}$ ${\cal O}(m_{c})$
  can be evaluated with the renormalization group (RG)
  equation \cite{9512380}
  \begin{equation}
  C_{1,2}({\mu}) = U_{4}({\mu},m_{b})U_{5}(m_{b},m_{W})C_{1,2}(m_{W})
  \label{ci},
  \end{equation}
  where $U_{f}({\mu}_{f},{\mu}_{i})$ is the RG evolution matrix
  which transforms the Wilson coefficients from scale of ${\mu}_{i}$
  to ${\mu}_{f}$.
  The expression for $U_{f}({\mu}_{f},{\mu}_{i})$
  can be found in Ref. \cite{9512380}.
  The numerical values of the leading-order (LO) and NLO
  $C_{1,2}$ in the naive dimensional regularization scheme
  are listed in Table \ref{tab:ai}.
  The values of coefficients $C_{1,2}$ in Table \ref{tab:ai}
  agree well with those obtained with
  ``effective'' number of active flavors $f$ $=$ $4.15$ \cite{9512380}
  rather than formula Eq.(\ref{ci}).

  To obtain the decay amplitudes and branching ratios,
  the remaining works are to evaluate accurately the
  hadronic matrix elements (HME) where the local operators are
  sandwiched between the charmonium and final states,
  which is also the most intricate work in dealing
  with the weak decay of heavy hadrons by now.

  \subsection{Hadronic matrix elements}
  \label{sec0202}
  Analogous to the exclusive processes with perturbative
  QCD theory proposed by Lepage and Brodsky \cite{prd22},
  the QCDF approach is developed by Beneke {\em et al.}
  \cite{qcdf1} to deal with HME based on the collinear
  factorization approximation and power counting rules
  in the heavy quark limit,
  and has been extensively used for $B$ meson decays.
  Using the QCDF master formula, HME of nonleptonic decays
  could be written as the convolution integrals of the
  process-dependent hard scattering kernels and
  universal light-cone distribution amplitudes (LCDA) of
  participating hadrons.

  The spectator quark is the heavy-flavor charm quark for
  charmonium weak decays into $DM$ final states.
  It is commonly assumed that the virtuality of the
  gluon connecting to the heavy spectator
  is of order ${\Lambda}_{\rm QCD}^{2}$, where
  ${\Lambda}_{\rm QCD}$ is the characteristic QCD scale.
  Hence, the transition form factors between
  charmonium and $D$ mesons
  are assumed to be dominated by the soft and nonperturbative
  contributions, and the amplitudes of the spectator
  rescattering subprocess are power-suppressed \cite{qcdf1}.
  Taking the ${\eta}_{c}$ ${\to}$ $DM$ decays for example,
  HME can be written as
   \begin{equation}
  {\langle}DM{\vert}Q_{1,2}{\vert}{\eta}_{c}{\rangle}
   \ =\
   \sum\limits_{i} F_{i}^{ {\eta}_{c}{\to}D }\,f_{M}\,
  {\int}\,dx\, H_{i}(x)\,{\Phi}_{M}(x)
   \label{hadronic},
   \end{equation}
  where $F_{i}^{ {\eta}_{c}{\to}D }$ is the weak transition form
  factor, $f_{M}$ and ${\Phi}_{M}(x)$ are the decay constant
  and LCDA of the meson $M$, respectively.
  The leading twist LCDA for the pseudoscalar and longitudinally
  polarized vector mesons can  be expressed in terms of Gegenbauer
  polynomials \cite{dap,dav}:
   \begin{equation}
  {\Phi}_{M}(x)=6\,x\bar{x} \sum\limits_{n=0}^{\infty}
   a_{n}^{M}\, C_{n}^{3/2}(x-\bar{x})
   \label{twist},
   \end{equation}
  where $\bar{x}$ $=$ $1$ $-$ $x$; $C_{n}^{3/2}(z)$ is
  the Gegenbauer polynomial,
   \begin{equation}
   C_{0}^{3/2}(z)= 1, \quad
   C_{1}^{3/2}(z)= 3\,z, \quad
   C_{2}^{3/2}(z)= \frac{3}{2}(5\,z^{2}-1), \quad
   {\cdots}
   \label{phi-cn},
   \end{equation}
  $a_{n}^{M}$ is the Gegenbauer moment corresponding to
  the Gegenbauer polynomials $C_{n}^{3/2}(z)$;
  $a_{0}^{M}$ ${\equiv}$ $1$ for the asymptotic form;
  $a_{n}$ $=$ $0$ for $n$ $=$ 1, 3, 5, ${\cdots}$
  because of the $G$-parity invariance of the ${\pi}$,
  ${\eta}^{(\prime)}$, ${\rho}$, ${\omega}$, ${\phi}$
  meson distribution amplitudes. In this paper, to give
  a rough estimation, the contributions from higher-order
  $n$ ${\ge}$ $3$ Gegenbauer polynomials are not considered
  for the moment.

  Hard scattering function $H_{i}(x)$ in Eq.(\ref{hadronic}) is,
  in principle, calculable order by order with the perturbative
  QCD theory.
  At the order of ${\alpha}_{s}^{0}$, $H_{i}(x)$ $=$ $1$.
  This is the simplest scenario, and one goes back to the
  naive factorization where there is no information about
  the strong phases and the renormalization scale hidden
  in the HME.
  At the order of ${\alpha}_{s}$ and higher orders,
  the renormalization scale dependence of hadronic matrix elements
  could be recuperated to partly cancel the ${\mu}$-dependence of
  the Wilson coefficients.
  In addition, part of the strong phases could be reproduced from
  nonfactorizable contributions.

  Within the QCDF framework, amplitudes for ${\eta}_{c}$
  ${\to}$ $DM$ decays can be expressed as:
   \begin{equation}
  {\cal A}({\eta}_{c}{\to}DM)\ =\
  {\langle}DM{\vert}{\cal H}_{\rm eff}{\vert}{\eta}_{c}{\rangle}\ =\
   \frac{G_{F}}{\sqrt{2}}\,
   V_{cq_{1}}^{\ast} V_{uq_{2}}\, a_{i}\,
  {\langle}M{\vert}J^{\mu}{\vert}0{\rangle}
  {\langle}D{\vert}J_{\mu}{\vert}{\eta}_{c}{\rangle}
   \label{lorentz}.
   \end{equation}

  In addition, the HME for the ${\psi}(1S,2S)$ ${\to}$ $DV$
  decays are conventionally expressed as the helicity
  amplitudes with the decomposition \cite{vv1,vv2},
    \begin{eqnarray}
   {\cal H}_{\lambda} & =&
   {\langle}V{\vert}J^{\mu}{\vert}0{\rangle}
   {\langle}D{\vert}J_{\mu}{\vert}{\psi}{\rangle}
    \nonumber \\ &=&
   {\epsilon}_{V}^{{\ast}{\mu}} {\epsilon}_{\psi}^{\nu} \Big\{
    a\,g_{{\mu}{\nu}}
    +\frac{ b  }{m_{\psi}\,m_{V}} ( p_{\psi} + p_{D} )^{\mu} p_{V}^{\nu}
    +\frac{i\,c }{m_{\psi}\,m_{V}} {\epsilon}_{{\mu}{\nu}{\alpha}{\beta}}
     p_{V}^{\alpha}(p_{\psi}+p_{D})^{\beta} \Big\}
    \label{spd}.
    \end{eqnarray}
  The relations among helicity amplitudes and invariant
  amplitudes $a$, $b$, $c$ are
    \begin{eqnarray}
   {\cal H}_{0} &=& -a\,x-2b\,(x^{2}-1)
    \label{h0}, \\
   {\cal H}_{\pm} &=& a\, {\pm}\,2c\,\sqrt{x^{2}-1}
    \label{h1},
    \end{eqnarray}
    \begin{equation}
    x \ =\ \frac{p_{\psi}{\cdot}p_{V}}{m_{\psi}\,m_{V}}
      \ =\ \frac{m_{\psi}^{2}-m_{D}^{2}+m_{V}^{2}}{2\,m_{\psi}\,m_{V}}
    \label{xx},
    \end{equation}
  where three scalar amplitudes $a$, $b$, $c$ describe the
  $s$, $d$, $p$ wave contributions, respectively.

  The effective coefficient $a_{i}$ at the order of
  ${\alpha}_{s}$ can be expressed as \cite{qcdf1}:
  \begin{eqnarray}
   a_{1}
   &=& C_{1}^{\rm NLO}+\frac{1}{N_{c}}\,C_{2}^{\rm NLO}
    + \frac{{\alpha}_{s}}{4{\pi}}\, \frac{C_{F}}{N_{c}}\,
      C_{2}^{\rm LO}\, {\cal V}
   \label{a1}, \\
   a_{2}
   &=& C_{2}^{\rm NLO}+\frac{1}{N_{c}}\,C_{1}^{\rm NLO}
    + \frac{{\alpha}_{s}}{4{\pi}}\, \frac{C_{F}}{N_{c}}\,
      C_{1}^{\rm LO}\, {\cal V}
   \label{a2},
  \end{eqnarray}
  where the color factor $C_{F}$ $=$ $4/3$;
  the color number $N_{c}$ $=$ $3$.
  For the transversely polarized light vector meson, the factor
  ${\cal V}$ $=$ $0$ in the helicity ${\cal H}_{\pm}$
  amplitudes beyond the leading twist contributions.
  With the leading twist LCDA for the pseudoscalar and
  longitudinally polarized vector mesons, the factor
  ${\cal V}$ is written as \cite{qcdf1}:
  \begin{equation}
  {\cal V} = 6\,{\log}\Big( \frac{m_{c}^{2}}{{\mu}^{2}} \Big)
    -  18 - \Big( \frac{1}{2}+i3{\pi} \Big)
    +  \Big( \frac{11}{2}-i3{\pi} \Big)\,a_{1}^{M}
    -   \frac{21}{20}\,a_{2}^{M} +{\cdots}
  \label{vc}.
  \end{equation}

  From the numbers in Table. \ref{tab:ai}, it is found that
  (1) the values of coefficients $a_{1,2}$ agree generally with those
  used in previous works \cite{epjc55,plb252,ijma14,adv2013,cheng}.
  (2) The strong phases appear by taking nonfactorizable corrections
  into account, which is necessary for $CP$ violation.
  (3) The strong phase of $a_{1}$ is small due to the
  suppression of ${\alpha}_{s}$ and $1/N_{c}$.
  The strong phase of $a_{2}$ is large due to the
  enhancement from the large Wilson coefficients $C_{1}$.

   \begin{table}[h]
   \caption{Numerical values of the Wilson coefficients $C_{1,2}$
    and parameters $a_{1,2}$ for the ${\eta}_{c}$
    ${\to}$ $D{\pi}$ decay with $m_{c}$ $=$ 1.275 GeV \cite{pdg},
    where $a_{1,2}$ in Ref. \cite{cheng} is used in the $D$ meson
    weak decay.}
   \label{tab:ai}
   \begin{ruledtabular}
   \begin{tabular}{c|cc|cc|cc||c|c|c}
 & \multicolumn{2}{c|}{LO}
 & \multicolumn{2}{c|}{NLO}
 & \multicolumn{2}{c||}{QCDF}
 & \multicolumn{3}{c}{Previous works} \\ \cline{2-10}
 ${\mu}$ & $C_{1}$ & $C_{2}$ & $C_{1}$ & $C_{2}$ & $a_{1}$ & $a_{2}$
         & Ref. & $a_{1}$ & $a_{2}$ \\ \hline
 $0.8\,m_{c}$ & $1.335$ & $-0.589$ & $1.275$ & $-0.504$
              & $1.275e^{+i4^{\circ}}$ & $0.503e^{-i154^{\circ}}$
              & \cite{epjc55,ijma14,adv2013} & $1.26$ & $-0.51$ \\
 $m_{c}$      & $1.276$ & $-0.505$ & $1.222$ & $-0.425$
              & $1.219e^{+i3^{\circ}}$ & $0.402e^{-i154^{\circ}}$
              & \cite{plb252} & $1.3{\pm}0.1$ & $-0.55{\pm}0.10$  \\
 $1.2\,m_{c}$ & $1.240$ & $-0.450$ & $1.190$ & $-0.374$
              & $1.186e^{+i3^{\circ}}$ & $0.342e^{-i154^{\circ}}$
              & \cite{cheng} & $1.274$ & $-0.529$
   \end{tabular}
   \end{ruledtabular}
   \end{table}

  \subsection{Form factors}
  \label{sec0203}
  The weak transition form factors between charmonium and
  a charmed meson are defined as follows \cite{bsw}:
    \begin{eqnarray}
   & &
   {\langle}D(p_{2}){\vert}V_{\mu}-A_{\mu}
   {\vert}{\eta}_{c}(p_{1}){\rangle}
    \nonumber \\ &=&
    \Big\{ (p_{1}+p_{2})_{\mu}
   -\frac{ m_{{\eta}_{c}}^{2} - m_{D}^{2} }{ q^{2} }\, q_{\mu}
    \Big\} F_{1}(q^{2})
   +\frac{ m_{{\eta}_{c}}^{2} - m_{D}^{2} }{ q^{2} }\, q_{\mu}
    F_{0}(q^{2})
    \label{f0f1},
    \end{eqnarray}
    \begin{eqnarray}
   & &
   {\langle}D(p_{2}){\vert}V_{\mu}-A_{\mu}
   {\vert}{\psi}(p_{1},{\epsilon}){\rangle}
    \nonumber \\ &=&
  -{\epsilon}_{{\mu}{\nu}{\alpha}{\beta}}\,
   {\epsilon}_{{\psi}}^{{\nu}}\,
    q^{\alpha}\, (p_{1}+p_{2})^{\beta}\,
     \frac{V(q^{2})}{m_{{\psi}}+m_{D}}
   -i\,\frac{2\,m_{{\psi}}\,{\epsilon}_{{\psi}}{\cdot}q}{q^{2}}\,
    q_{\mu}\, A_{0}(q^{2})
    \nonumber \\ & &
    -i\,{\epsilon}_{{\psi},{\mu}}\,
    ( m_{\psi}+m_{D} )\, A_{1}(q^{2})
   -i\,\frac{{\epsilon}_{\psi}{\cdot}q}{m_{{\psi}}+m_{D}}\,
   ( p_{1} + p_{2} )_{\mu}\, A_{2}(q^{2})
    \nonumber \\ & &
   +i\,\frac{2\,m_{\psi}\,{\epsilon}_{\psi}{\cdot}q}{q^{2}}\,
   q_{\mu}\, A_{3}(q^{2})
    \label{a0123v},
    \end{eqnarray}
  where $q$ $=$ $p_{1}$ $-$ $p_{2}$;
  ${\epsilon}_{\psi}$ denotes the ${\psi}$'s polarization vector.
  The form factors
  $F_{0}(0)$ $=$ $F_{1}(0)$ and $A_{0}(0)$ $=$ $A_{3}(0)$
  are required compulsorily to cancel singularities at the
  pole of $q^{2}$ $=$ $0$.
  There is a relation among these form factors
  \begin{equation}
  2m_{\psi}A_{3}(q^{2})=(m_{\psi}+m_{D})A_{1}(q^{2})+(m_{\psi}-m_{D})A_{2}(q^{2})
  \label{form01}.
  \end{equation}

  There are four independent transition form factors,
  $F_{0}(0)$, $A_{0,1}(0)$ and $V(0)$, at the pole of
  $q^{2}$ $=$ $0$. They could be written as the overlap
  integrals of wave functions \cite{bsw}.
  \begin{equation}
  F_{0}(0) =
  {\int}d\vec{k}_{\perp} {\int}_{0}^{1}dx\,
   \Big\{ {\Phi}_{{\eta}_{c}}(\vec{k}_{\perp},x,0,0)\,
  {\Phi}_{D}(\vec{k}_{\perp},x,0,0) \Big\}
  \label{form-f0},
  \end{equation}
  \begin{equation}
  A_{0}(0) =
  {\int}d\vec{k}_{\perp} {\int}_{0}^{1}dx\,
   \Big\{ {\Phi}_{\psi}(\vec{k}_{\perp},x,1,0)\,
  {\sigma}_{z}\, {\Phi}_{D}(\vec{k}_{\perp},x,0,0) \Big\}
  \label{form-a0},
  \end{equation}
  \begin{eqnarray}
  A_{1}(0) &=& \frac{ m_{c}+m_{q} }{ m_{\psi}+m_{D} } I
  \label{form-a1}, \\
  V(0) &=& \frac{ m_{c}-m_{q} }{ m_{\psi}-m_{D} } I
  \label{form-v},
  \end{eqnarray}
  \begin{equation}
  I = \sqrt{2}
  {\int}d\vec{k}_{\perp} {\int}_{0}^{1} \frac{dx}{x}\,
   \Big\{ {\Phi}_{\psi}(\vec{k}_{\perp},x,1,-1)\,
  i{\sigma}_{y}\, {\Phi}_{D}(\vec{k}_{\perp},x,0,0) \Big\}
  \label{form-ii},
  \end{equation}
  where ${\sigma}_{y,z}$ is the Pauli matrix acting on
  the spin indices of the decaying charm quark;
  $x$ and $\vec{k}_{\perp}$ denote the fraction of
  the longitudinal momentum and the transverse momentum
  of the nonspectator quark, respectively.

  With the separation of the spin and spacial variables,
  wave functions can be written as
  \begin{equation}
  {\Phi}(\vec{k}_{\perp},x,j,j_{z})\, =
  {\phi}(\vec{k}_{\perp},x)\,{\vert}s,s_{z},s_{1},s_{2}{\rangle}
  \label{wave01},
  \end{equation}
  where the total angular momentum $\vec{j}$ $=$ $\vec{L}$
  $+$ $\vec{s}_{1}$ $+$ $\vec{s}_{2}$ $=$ $\vec{s}_{1}$
  $+$ $\vec{s}_{2}$ $=$ $\vec{s}$
  because the orbital angular momentum between the valence
  quarks in the ${\psi}(1S,2S)$, ${\eta}_{c}(1S,2S)$,
  $D$ mesons in question have $\vec{L}$ $=$ $0$;
  $s_{1,2}$ denote the spins of valence quarks in meson;
  $s$ $=$ $1$ and $0$ for the ${\psi}$ and
  ${\eta}_{c}$ mesons, respectively.

  The charm quark in the charmonium state is nearly nonrelativistic with
  an average velocity $v$ ${\ll}$ $1$  
  based on arguments of nonrelativistic quantum chromodynamics
  (NRQCD) \cite{prd46,prd51,rmp77}.
  For the $D$ meson, the valence quarks are also
  nonrelativistic due to $m_{D}$ ${\approx}$ $m_{c}$ $+$
  $m_{q}$, where the light quark mass $m_{u}$ ${\approx}$
  $m_{d}$ ${\approx}$ 310 MeV and $m_{s}$ ${\approx}$ 510
  MeV \cite{kamal}.
  Here, we will take the solution of the Schr\"{o}dinger equation
  with a scalar harmonic oscillator potential as the wave functions
  of the charmonium and $D$ mesons :
   \begin{equation}
  {\phi}_{1S}(\vec{k})\ {\sim}\
   e^{-\vec{k}^{2}/2{\alpha}^{2}}
   \label{wave-r1s},
   \end{equation}
   \begin{equation}
  {\phi}_{2S}(\vec{k})\ {\sim}\
   e^{-\vec{k}^{2}/2{\alpha}^{2}}
   ( 2\vec{k}^{2}-3{\alpha}^{2} )
   \label{wave-r2s},
   \end{equation}
  where the parameter ${\alpha}$ determines the average
  transverse quark momentum,
  ${\langle}{\phi}_{1S}{\vert}\vec{k}^{2}_{\perp}{\vert}{\phi}_{1S}{\rangle}$
  $=$ ${\alpha}^{2}$.
  With the NRQCD power counting rules \cite{prd46},
  ${\vert}\vec{k}_{\perp}{\vert}$ ${\sim}$ $mv$ ${\sim}$
  $m{\alpha}_{s}$ for heavy quarkonium.
  Hence, parameter ${\alpha}$ is approximately taken as
  $m{\alpha}_{s}$ in our calculation.

  Using the substitution ansatz \cite{xiao},
   \begin{equation}
   \vec{k}^{2}\ {\to}\
   \frac{\vec{k}_{\perp}^{2}+\bar{x}\,m_{q}^{2}+x\,m_{c}^{2}}{4\,x\,\bar{x}}
   \label{wave04},
   \end{equation}
   one can obtain
   \begin{equation}
  {\phi}_{1S}(\vec{k}_{\perp},x) = A\,
  {\exp}\Big\{ \frac{\vec{k}_{\perp}^{2}+\bar{x}\,m_{q}^{2}+x\,m_{c}^{2}}
                    {-8\,{\alpha}^{2}\,x\,\bar{x}} \Big\}
   \label{wave-k1s},
   \end{equation}
   \begin{equation}
  {\phi}_{2S}(\vec{k}_{\perp},x) = B\, {\phi}_{1S}(\vec{k}_{\perp},x)
   \Big\{ \frac{\vec{k}_{\perp}^{2}+\bar{x}\,m_{q}^{2}+x\,m_{c}^{2}}
                    {6\,{\alpha}^{2}\,x\,\bar{x}}-1 \Big\}
   \label{wave-k2s},
   \end{equation}
   where 
   the parameters $A$ and $B$ are the normalization coefficients
   satisfying with the normalization condition,
  \begin{equation}
  {\int}d\vec{k}_{\perp}
  {\int}_{0}^{1}dx\, {\vert}{\phi}(\vec{k}_{\perp},x)
  {\vert}^{2}\, = 1
  \label{wave02}.
  \end{equation}

   \begin{table}[h]
   \caption{The numerical values of transition form factors
   at $q^{2}$ $=$ $0$, where uncertainties of this work
   come from the charm quark mass.}
   \label{tab:ff}
   \begin{ruledtabular}
   \begin{tabular}{l|c|cccc}
   Transition & Reference &
   $F_{0}(0)$ & $A_{0}(0)$ & $A_{1}(0)$ & $V(0)$ \\ \hline
   ${\eta}_{c}(1S)$, ${\psi}(1S)$ ${\to}$ $D_{u,d}$
 & This work
 & $0.85{\pm}0.01$
 & $0.85{\pm}0.01$
 & $0.72{\pm}0.01$
 & $1.76{\pm}0.03$ \\
 & Ref. \cite{prd78}\footnotemark[1]
 & ...
 & $0.68{\pm}0.01$
 & $0.68{\pm}0.01$
 & $1.6{\pm}0.1$ \\
 & Ref. \cite{epjc54}\footnotemark[2]
 & ...
 & $0.27^{+0.02}_{-0.03}$
 & $0.27^{+0.03}_{-0.02}$
 & $0.81^{+0.12}_{-0.08}$ \\
 & Ref. \cite{plb252}\footnotemark[3]
 & ...
 & 0.40 (0.61)
 & 0.44 (0.68)
 & 1.17 (1.82) \\
 & Ref. \cite{adv2013}\footnotemark[4]
 & ...
 & $0.55{\pm}0.02$
 & $0.77^{+0.09}_{-0.07}$
 & $2.14^{+0.15}_{-0.11}$ \\
 & Ref. \cite{adv2013}\footnotemark[5]
 & ...
 & 0.54
 & 0.80
 & 2.21 \\ \hline
   ${\eta}_{c}(1S)$, ${\psi}(1S)$ ${\to}$ $D_{s}$
 & This work
 & $0.90{\pm}0.01$
 & $0.90{\pm}0.01$
 & $0.81{\pm}0.01$
 & $1.55{\pm}0.04$ \\
 & Ref. \cite{prd78}\footnotemark[1]
 & ...
 & $0.68{\pm}0.01$
 & $0.68{\pm}0.01$
 & $1.8$ \\
 & Ref. \cite{epjc54}\footnotemark[2]
 & ...
 & $0.37{\pm}0.02$
 & $0.38^{+0.02}_{-0.01}$
 & $1.07^{+0.05}_{-0.02}$ \\
 & Ref. \cite{plb252}\footnotemark[3]
 & ...
 & 0.47 (0.66)
 & 0.55 (0.78)
 & 1.25 (1.80) \\
 & Ref. \cite{adv2013}\footnotemark[4]
 & ...
 & $0.71^{+0.04}_{-0.02}$
 & $0.94{\pm}0.07$
 & $2.30^{+0.09}_{-0.06}$ \\
 & Ref. \cite{adv2013}\footnotemark[5]
 & ...
 & 0.69
 & 0.96
 & 2.36 \\ \hline
   ${\eta}_{c}(2S)$, ${\psi}(2S)$ ${\to}$ $D_{u,d}$
 & This work
 & $0.62{\pm}0.01$
 & $0.61{\pm}0.01$
 & $0.54{\pm}0.01$
 & $1.00{\pm}0.04$ \\
   ${\eta}_{c}(2S)$, ${\psi}(2S)$ ${\to}$ $D_{s}$
 & This work
 & $0.65{\pm}0.01$
 & $0.64{\pm}0.01$
 & $0.59{\pm}0.02$
 & $0.83{\pm}0.04$
   \end{tabular}
   \end{ruledtabular}
   \footnotetext[1]{The form factors are computed
   with the covariant light-front quark model,
   where uncertainties come from the decay constant of charmed meson.}
   \footnotetext[2]{The form factors are computed
   with QCD sum rules, where uncertainties
   are from the Borel parameters.}
   \footnotetext[3]{The form factors are computed
   with parameter ${\omega}$ $=$ 0.4 (0.5) GeV
   using the WSB model.}
   \footnotetext[4]{The form factors are computed
   with flavor dependent parameter ${\omega}$
   using the WSB model.}
   \footnotetext[5]{The form factors are computed
   with parameter ${\omega}$ $=$ $m{\alpha}_{s}$
   using the WSB model.}
   \end{table}

   The numerical values of transition form factors
   at $q^{2}$ $=$ $0$ are listed in Table \ref{tab:ff}.
   It is found that
   (1) the model dependence of form factors is large;
   (2) isospin-breaking effects are negligible
   and flavor breaking effects are small;
   (3) as stated in Ref. \cite{bsw} that
   $F_{0}$ ${\simeq}$ $A_{0}$ holds within collinear symmetry.

  \section{Numerical results and discussion}
  \label{sec03}
  In the charmonium center-of-mass frame, the branching ratio
  for the charmonium weak decay can be written as
   \begin{equation}
  {\cal B}r({\eta}_{c}{\to}DM)\ =\
   \frac{ p_{\rm cm} }{ 4\,{\pi}\,m_{{\eta}_{c}}^{2}\, {\Gamma}_{{\eta}_{c}} }\,
  {\vert}{\cal A}({\eta}_{c}{\to}DM){\vert}^{2}
   \label{br01},
   \end{equation}
   \begin{equation}
  {\cal B}r({\psi}{\to}DM)\ =\
   \frac{ p_{\rm cm} }{ 12\,{\pi}\,m_{\psi}^{2}\, {\Gamma}_{\psi} }\,
  {\vert}{\cal A}({\psi}{\to}DM){\vert}^{2}
   \label{br02},
   \end{equation}
  where the common momentum of final states is
   \begin{equation}
   p_{\rm cm}\ =\
   \frac{ \sqrt{ [m_{{\eta}_{c},{\psi}}^{2}-(m_{D}+m_{M})^{2}]
                 [m_{{\eta}_{c},{\psi}}^{2}-(m_{D}-m_{M})^{2}] }  }
        { 2\,m_{{\eta}_{c},{\psi}} }
   \label{pcm};
   \end{equation}
  The decay amplitudes for ${\cal A}({\psi}{\to}DM)$
  and ${\cal A}({\eta}_{c}{\to}DM)$ are collected in
  Appendix \ref{app01} and \ref{app02}, respectively.

  In our calculation, we assume that the light vector mesons
  are ideally mixed, i.e.,
  ${\omega}$ $=$ $(u\bar{u}+d\bar{d})/\sqrt{2}$
  and ${\phi}$ $=$ $s\bar{s}$.
  For the mixing of pseudoscalar ${\eta}$ and ${\eta}^{\prime}$
  meson, we will adopt the quark-flavor basis description proposed
  in \cite{eta}, and neglect the contributions from
  possible gluonium compositions, i.e.,
   \begin{equation}
   \left(\begin{array}{c}
  {\eta} \\ {\eta}^{\prime}
   \end{array}\right) =
   \left(\begin{array}{cc}
  {\cos}{\phi} & -{\sin}{\phi} \\
  {\sin}{\phi} &  {\cos}{\phi}
   \end{array}\right)
   \left(\begin{array}{c}
  {\eta}_{q} \\ {\eta}_{s}
   \end{array}\right)
   \label{mixing01},
   \end{equation}
  where ${\eta}_{q}$ $=$ $(u\bar{u}+d\bar{d})/{\sqrt{2}}$
  and ${\eta}_{s}$ $=$ $s\bar{s}$; the mixing angle
  ${\phi}$ $=$ $(39.3{\pm}1.0)^{\circ}$ \cite{eta}.
  The mass relations are
   \begin{eqnarray}
   m_{{\eta}_{q}}^{2}&=& \displaystyle
   m_{\eta}^{2}{\cos}^{2}{\phi}
  +m_{{\eta}^{\prime}}^{2}{\sin}^{2}{\phi}
  -\frac{\sqrt{2}f_{{\eta}_{s}}}{f_{{\eta}_{q}}}
  (m_{{\eta}^{\prime}}^{2}- m_{\eta}^{2})\,
  {\cos}{\phi}\,{\sin}{\phi}
   \label{ss12}, \\
   m_{{\eta}_{s}}^{2}&=& \displaystyle
   m_{\eta}^{2}{\sin}^{2}{\phi}
  +m_{{\eta}^{\prime}}^{2}{\cos}^{2}{\phi}
  -\frac{f_{{\eta}_{q}}}{\sqrt{2}f_{{\eta}_{s}}}
  (m_{{\eta}^{\prime}}^{2}- m_{\eta}^{2})\,
  {\cos}{\phi}\,{\sin}{\phi}
   \label{ss13}.
   \end{eqnarray}

  \begin{table}[h]
  \caption{Numerical values of input parameters.}
   \label{tab:in}
  \begin{ruledtabular}
  \begin{tabular}{ll}
    ${\lambda}$ $=$ $0.22537{\pm}0.00061$      \cite{pdg}
  & $A$         $=$ $0.814^{+0.023}_{-0.024}$  \cite{pdg} \\
    $\bar{\rho}$ $=$ $0.117{\pm}0.021$         \cite{pdg}
  & $\bar{\eta}$ $=$ $0.353{\pm}0.013$         \cite{pdg} \\ \hline
    $m_{c}$ $=$ $1.275{\pm}0.025$ GeV  \cite{pdg}
  & $m_{D_{u}}$ $=$ $1864.84{\pm}0.07$ MeV \cite{pdg} \\
    $m_{D_{d}}$ $=$ $1869.61{\pm}0.10$ MeV \cite{pdg}
  & $m_{D_{s}}$ $=$ $1968.30{\pm}0.11$ MeV \cite{pdg} \\ \hline
    $f_{\pi}$      $=$ $130.41{\pm}0.20$ MeV \cite{pdg}
  & $f_{K}  $      $=$ $156.2{\pm}0.7$ MeV \cite{pdg} \\
    $f_{{\eta}_{q}}$ $=$ $(1.07{\pm}0.02)\,f_{\pi}$ \cite{eta}
  & $f_{{\eta}_{s}}$ $=$ $(1.34{\pm}0.06)\,f_{\pi}$ \cite{eta} \\
    $f_{\rho}$   $=$ $216{\pm}3$ MeV \cite{dav}
  & $f_{\omega}$ $=$ $187{\pm}5$ MeV \cite{dav} \\
    $f_{\phi}$   $=$ $215{\pm}5$ MeV \cite{dav}
  & $f_{K^{\ast}}$ $=$ $220{\pm}5$ MeV \cite{dav} \\ \hline
    $a_{2}^{\pi}$ $=$ $a_{2}^{{\eta}_{q}}$ $=$
    $a_{2}^{{\eta}_{s}}$ $=$ $0.25{\pm}0.15$ \cite{dap}
  & $a_{2}^{\rho}$ $=$ $a_{2}^{\omega}$ $=$ $0.15{\pm}0.07$ \cite{dav} \\
    $a_{1}^{\bar{K}}$ $=$ $-a_{1}^{K}$ $=$ $0.06{\pm}0.03$ \cite{dap}
  & $a_{2}^{K}$ $=$ $a_{2}^{\bar{K}}$ $=$ $0.25{\pm}0.15$ \cite{dap} \\
    $a_{1}^{\bar{K}^{\ast}}$ $=$ $-a_{1}^{K^{\ast}}$ $=$ $0.03{\pm}0.02$ \cite{dav}
  & $a_{2}^{K^{\ast}}$ $=$ $a_{2}^{\bar{K}^{\ast}}$ $=$ $0.11{\pm}0.09$ \cite{dav} \\
    $a_{1}^{\pi}$ $=$ $a_{1}^{\rho}$ $=$ $a_{1}^{\omega}$ $=$ $a_{1}^{\phi}$ $=$ $0$
  & $a_{2}^{\phi}$ $=$ $0.18{\pm}0.08$ \cite{dav}
  \end{tabular}
  \end{ruledtabular}
  \end{table}

  {\renewcommand{\baselinestretch}{1.3}
   \begin{sidewaystable}[h]
   \caption{Branching ratios for the nonleptonic two-body $J/{\psi}$ weak
   decays, where the uncertainties of this work come from the CKM parameters,
   the renormalization scale ${\mu}$ $=$ $(1{\pm}0.2)m_{c}$, hadronic
   parameters including decay constants and Gegenbauer moments, respectively.
   The results of Refs. \cite{epjc55,adv2013,ijma14} are calculated with
   $a_{1}$ = 1.26 and $a_{2}$ $=$ $-0.51$.
   The results of Ref. \cite{epjc55} are based on QCD sum rules.
   The numbers in columns of ``A'', ``B'', ``C'' and ``D'' are based
   on the WSB model with flavor dependent ${\omega}$, QCD inspired
   ${\omega}$ $=$ ${\alpha}_{s}m$, universal
   ${\omega}$ $=$ 0.4 GeV and 0.5 GeV, respectively.}
   \label{tab:psi}
   \begin{ruledtabular}
   \begin{tabular}{lccccccc}
   final & & Ref. \cite{epjc55}
           & \multicolumn{3}{c}{ Ref. \cite{adv2013} }
           & Ref. \cite{ijma14}
           & This \\ \cline{4-6}
   states  & case & & A & B & C & D & work \\ \hline
 $D_{s}^{-}{\pi}^{+}$
  & 1-a
  & $2.0 {\times}10^{-10}$
  & $7.41{\times}10^{-10}$
  & $7.13{\times}10^{-10}$
  & $3.32{\times}10^{-10}$
  & $8.74{\times}10^{-10}$
  & ($1.09^{+0.01+0.10+0.01}_{-0.01-0.06-0.01}){\times}10^{-9}$
 \\
 $D_{s}^{-}K^{+}$
  & 1-b
  & $1.6{\times}10^{-11}$
  & $5.3{\times}10^{-11}$
  & $5.2{\times}10^{-11}$
  & $2.4{\times}10^{-11}$
  & $5.5{\times}10^{-11}$
  & ($6.18^{+0.03+0.59+0.08}_{-0.03-0.33-0.08}){\times}10^{-11}$
 \\
 $D_{d}^{-}{\pi}^{+}$
  & 1-b
  & $0.8{\times}10^{-11}$
  & $2.9{\times}10^{-11}$
  & $2.8{\times}10^{-11}$
  & $1.5{\times}10^{-11}$
  & $5.5{\times}10^{-11}$
  & ($6.37^{+0.03+0.60+0.03}_{-0.03-0.34-0.03}){\times}10^{-11}$
 \\
 $D_{d}^{-}K^{+}$
  & 1-c
  & ...
  & $2.3{\times}10^{-12}$
  & $2.2{\times}10^{-12}$
  & $1.2{\times}10^{-12}$
  & ...
  & ($3.79^{+0.04+0.36+0.05}_{-0.04-0.20-0.05}){\times}10^{-12}$
 \\
 $\overline{D}_{u}^{0}{\pi}^{0}$
  & 2-b
  & ...
  & $2.4{\times}10^{-12}$
  & $2.3{\times}10^{-12}$
  & $1.2{\times}10^{-12}$
  & $5.5{\times}10^{-12}$
  & ($3.50^{+0.02+1.98+0.06}_{-0.02-0.97-0.06}){\times}10^{-12}$
 \\
 $\overline{D}_{u}^{0}K^{0}$
  & 2-c
  & ...
  & $4.0{\times}10^{-13}$
  & $4.0{\times}10^{-13}$
  & $2.0{\times}10^{-13}$
  & ...
  & ($4.16^{+0.04+2.35+0.11}_{-0.04-1.15-0.10}){\times}10^{-13}$
 \\
 $\overline{D}_{u}^{0}\overline{K}^{0}$
  & 2-a
  & $3.6{\times}10^{-11}$
  & $1.39{\times}10^{-10}$
  & $1.34{\times}10^{-10}$
  & $7.2{\times}10^{-11}$
  & $2.8{\times}10^{-10}$
  & ($1.44^{+0.01+0.81+0.03}_{-0.01-0.40-0.03}){\times}10^{-10}$
 \\
 $\overline{D}_{u}^{0}{\eta}$
  &
  & ...
  & $7.0{\times}10^{-12}$
  & $6.7{\times}10^{-12}$
  & $3.6{\times}10^{-12}$
  & $1.6{\times}10^{-12}$
  & ($1.03^{+0.01+0.58+0.10}_{-0.01-0.28-0.10}){\times}10^{-11}$
 \\
 $\overline{D}_{u}^{0}{\eta}^{\prime}$
  &
  & ...
  & $4.0{\times}10^{-13}$
  & $4.0{\times}10^{-13}$
  & $2.0{\times}10^{-13}$
  & $3.0{\times}10^{-13}$
  & ($5.83^{+0.03+3.29+1.72}_{-0.03-1.61-1.50}){\times}10^{-13}$
 \\ \hline
 $D_{s}^{-}{\rho}^{+}$
  & 1-a
  & $1.26{\times}10^{-9}$
  & $5.11{\times}10^{-9}$
  & $5.32{\times}10^{-9}$
  & $1.77{\times}10^{-9}$
  & $3.63{\times}10^{-9}$
  & ($3.82^{+0.01+0.36+0.11}_{-0.01-0.20-0.11}){\times}10^{-9}$
 \\
 $D_{s}^{-}K^{{\ast}+}$
  & 1-b
  & $0.82{\times}10^{-10}$
  & $2.82{\times}10^{-10}$
  & $2.96{\times}10^{-10}$
  & $0.97{\times}10^{-10}$
  & $2.12{\times}10^{-10}$
  & ($2.00^{+0.01+0.19+0.10}_{-0.01-0.11-0.09}){\times}10^{-10}$
 \\
 $D_{d}^{-}{\rho}^{+}$
  & 1-b
  & $0.42{\times}10^{-10}$
  & $2.16{\times}10^{-10}$
  & $2.28{\times}10^{-10}$
  & $0.72{\times}10^{-10}$
  & $2.20{\times}10^{-10}$
  & ($2.12^{+0.01+0.20+0.06}_{-0.01-0.11-0.06}){\times}10^{-10}$
 \\
 $D_{d}^{-}K^{{\ast}+}$
  & 1-c
  & ...
  & $1.3{\times}10^{-11}$
  & $1.3{\times}10^{-11}$
  & $4.2{\times}10^{-12}$
  & ...
  & ($1.14^{+0.01+0.11+0.06}_{-0.01-0.06-0.05}){\times}10^{-11}$
 \\
 $\overline{D}_{u}^{0}{\rho}^{0}$
  & 2-b
  & ...
  & $1.8{\times}10^{-11}$
  & $1.9{\times}10^{-11}$
  & $6.0{\times}10^{-12}$
  & $2.2{\times}10^{-11}$
  & ($1.08^{+0.01+0.61+0.04}_{-0.01-0.30-0.04}){\times}10^{-11}$
 \\
 $\overline{D}_{u}^{0}{\omega}$
  & 2-b
  & ...
  & $1.6{\times}10^{-11}$
  & $1.7{\times}10^{-11}$
  & $5.0{\times}10^{-12}$
  & $1.8{\times}10^{-11}$
  & ($8.10^{+0.04+4.56+0.50}_{-0.04-2.25-0.48}){\times}10^{-12}$
 \\
 $\overline{D}_{u}^{0}{\phi}$
  & 2-b
  & ...
  & $4.2{\times}10^{-11}$
  & $4.4{\times}10^{-11}$
  & $1.4{\times}10^{-11}$
  & $6.5{\times}10^{-11}$
  & ($1.92^{+0.01+1.08+0.10}_{-0.01-0.53-0.10}){\times}10^{-11}$
 \\
 $\overline{D}_{u}^{0}K^{{\ast}0}$
  & 2-c
  & ...
  & $2.1{\times}10^{-12}$
  & $2.2{\times}10^{-12}$
  & $7.0{\times}10^{-13}$
  & ...
  & ($1.19^{+0.01+0.67+0.07}_{-0.01-0.33-0.07}){\times}10^{-12}$
 \\
 $\overline{D}_{u}^{0}\overline{K}^{{\ast}0}$
  & 2-a
  & $1.54{\times}10^{-10}$
  & $7.61{\times}10^{-10}$
  & $8.12{\times}10^{-10}$
  & $2.51{\times}10^{-10}$
  & $1.03{\times}10^{-9}$
  & ($4.09^{+0.01+2.30+0.24}_{-0.01-1.14-0.23}){\times}10^{-10}$
  \end{tabular}
  \end{ruledtabular}
  \end{sidewaystable}
  }

   \begin{sidewaystable}[h]
   \caption{Branching ratios for the nonleptonic two-body ${\psi}(2S)$,
   ${\eta}_{c}(1S)$ and ${\eta}_{c}(2S)$ weak decays, where the uncertainties
   come from the CKM parameters, the renormalization scale ${\mu}$ $=$
   $(1{\pm}0.2)m_{c}$, hadronic parameters including decay
   constants and Gegenbauer moments, respectively.}
   \label{tab:etac}
   \begin{ruledtabular}
   \begin{tabular}{clccc}
   case & final states & ${\psi}(2S)$ decay
 & ${\eta}_{c}(1S)$ decay & ${\eta}_{c}(2S)$ decay \\ \hline
   1-a
 & $D_{s}^{-}{\pi}^{+}$
 & ($5.07^{+0.01+0.48+0.03}_{-0.01-0.27-0.02}){\times}10^{-10}$
 & ($7.35^{+0.01+0.69+0.04}_{-0.01-0.39-0.04}){\times}10^{-12}$
 & ($3.90^{+0.01+0.37+0.02}_{-0.01-0.21-0.02}){\times}10^{-11}$ \\
   1-b
 & $D_{s}^{-}K^{+}$
 & ($3.43^{+0.02+0.33+0.04}_{-0.02-0.18-0.04}){\times}10^{-11}$
 & ($4.97^{+0.03+0.48+0.06}_{-0.03-0.27-0.06}){\times}10^{-13}$
 & ($2.87^{+0.01+0.27+0.04}_{-0.01-0.15-0.04}){\times}10^{-12}$ \\
   1-b
 & $D_{d}^{-}{\pi}^{+}$
 & ($2.76^{+0.01+0.26+0.01}_{-0.01-0.15-0.01}){\times}10^{-11}$
 & ($4.39^{+0.02+0.41+0.02}_{-0.02-0.23-0.02}){\times}10^{-13}$
 & ($2.13^{+0.01+0.20+0.01}_{-0.01-0.11-0.01}){\times}10^{-12}$ \\
   1-c
 & $D_{d}^{-}K^{+}$
 & ($1.90^{+0.02+0.18+0.02}_{-0.02-0.10-0.02}){\times}10^{-12}$
 & ($3.04^{+0.03+0.29+0.04}_{-0.03-0.16-0.04}){\times}10^{-14}$
 & ($1.58^{+0.02+0.15+0.02}_{-0.02-0.08-0.02}){\times}10^{-13}$ \\
   2-b
 & $\overline{D}_{u}^{0}{\pi}^{0}$
 & ($1.51^{+0.01+0.85+0.02}_{-0.01-0.42-0.02}){\times}10^{-12}$
 & ($2.41^{+0.01+1.36+0.04}_{-0.01-0.67-0.04}){\times}10^{-14}$
 & ($1.16^{+0.01+0.66+0.02}_{-0.01-0.32-0.02}){\times}10^{-13}$ \\
   2-c
 & $\overline{D}_{u}^{0}K^{0}$
 & ($2.07^{+0.02+1.17+0.05}_{-0.02-0.57-0.05}){\times}10^{-13}$
 & ($3.35^{+0.04+1.89+0.09}_{-0.04-0.93-0.08}){\times}10^{-15}$
 & ($1.73^{+0.02+0.97+0.04}_{-0.02-0.48-0.04}){\times}10^{-14}$ \\
   2-a
 & $\overline{D}_{u}^{0}\overline{K}^{0}$
 & ($7.15^{+0.01+4.04+0.17}_{-0.01-1.98-0.16}){\times}10^{-11}$
 & ($1.16^{+0.01+0.65+0.03}_{-0.01-0.32-0.03}){\times}10^{-12}$
 & ($5.96^{+0.01+3.37+0.14}_{-0.01-1.65-0.14}){\times}10^{-12}$ \\
 & $\overline{D}_{u}^{0}{\eta}$
 & ($5.35^{+0.03+3.02+0.54}_{-0.03-1.48-0.50}){\times}10^{-12}$
 & ($8.66^{+0.04+4.89+0.88}_{-0.04-2.40-0.82}){\times}10^{-14}$
 & ($4.55^{+0.02+2.57+0.46}_{-0.02-1.26-0.43}){\times}10^{-13}$ \\
 & $\overline{D}_{u}^{0}{\eta}^{\prime}$
 & ($5.63^{+0.03+3.18+1.68}_{-0.03-1.56-1.46}){\times}10^{-13}$
 & ($7.66^{+0.04+4.32+2.28}_{-0.04-2.12-1.98}){\times}10^{-15}$
 & ($6.02^{+0.03+3.40+1.79}_{-0.03-1.67-1.56}){\times}10^{-14}$ \\  \hline
   1-a
 & $D_{s}^{-}{\rho}^{+}$
 & ($1.67^{+0.01+0.15+0.05}_{-0.01-0.09-0.05}){\times}10^{-9}$
 & ($5.28^{+0.01+0.50+0.15}_{-0.01-0.28-0.15}){\times}10^{-12}$
 & ($7.24^{+0.01+0.68+0.21}_{-0.01-0.38-0.21}){\times}10^{-11}$ \\
   1-b
 & $D_{s}^{-}K^{{\ast}+}$
 & ($9.59^{+0.05+0.89+0.46}_{-0.05-0.50-0.45}){\times}10^{-11}$
 & ($1.18^{+0.01+0.11+0.06}_{-0.01-0.06-0.06}){\times}10^{-13}$
 & ($3.47^{+0.02+0.33+0.17}_{-0.02-0.18-0.16}){\times}10^{-12}$ \\
   1-b
 & $D_{d}^{-}{\rho}^{+}$
 & ($8.99^{+0.05+0.83+0.26}_{-0.05-0.47-0.26}){\times}10^{-11}$
 & ($4.32^{+0.02+0.41+0.12}_{-0.02-0.23-0.12}){\times}10^{-13}$
 & ($4.13^{+0.02+0.39+0.12}_{-0.02-0.22-0.12}){\times}10^{-12}$ \\
   1-c
 & $D_{d}^{-}K^{{\ast}+}$
 & ($5.15^{+0.06+0.48+0.25}_{-0.05-0.27-0.24}){\times}10^{-12}$
 & ($1.38^{+0.01+0.13+0.07}_{-0.01-0.07-0.07}){\times}10^{-14}$
 & ($2.02^{+0.02+0.19+0.10}_{-0.02-0.11-0.10}){\times}10^{-13}$ \\
   2-b
 & $\overline{D}_{u}^{0}{\rho}^{0}$
 & ($4.36^{+0.02+2.44+0.15}_{-0.02-1.21-0.15}){\times}10^{-12}$
 & ($2.38^{+0.01+1.35+0.08}_{-0.01-0.66-0.08}){\times}10^{-14}$
 & ($2.24^{+0.01+1.27+0.08}_{-0.01-0.62-0.08}){\times}10^{-13}$ \\
   2-b
 & $\overline{D}_{u}^{0}{\omega}$
 & ($3.28^{+0.02+1.84+0.20}_{-0.02-0.91-0.19}){\times}10^{-12}$
 & ($1.74^{+0.01+0.98+0.11}_{-0.01-0.48-0.10}){\times}10^{-14}$
 & ($1.67^{+0.01+0.94+0.10}_{-0.01-0.46-0.10}){\times}10^{-13}$ \\
   2-b
 & $\overline{D}_{u}^{0}{\phi}$
 & ($9.40^{+0.05+5.28+0.52}_{-0.05-2.61-0.50}){\times}10^{-12}$
 & ($8.57^{+0.04+4.84+0.47}_{-0.04-2.38-0.45}){\times}10^{-15}$
 & ($3.28^{+0.02+1.85+0.18}_{-0.02-0.91-0.17}){\times}10^{-13}$ \\
   2-c
 & $\overline{D}_{u}^{0}K^{{\ast}0}$
 & ($5.09^{+0.05+2.86+0.31}_{-0.05-1.42-0.30}){\times}10^{-13}$
 & ($1.50^{+0.02+0.85+0.08}_{-0.02-0.42-0.08}){\times}10^{-15}$
 & ($2.18^{+0.02+1.23+0.12}_{-0.02-0.60-0.12}){\times}10^{-14}$ \\
   2-a
 & $\overline{D}_{u}^{0}\overline{K}^{{\ast}0}$
 & ($1.74^{+0.01+0.98+0.11}_{-0.01-0.49-0.10}){\times}10^{-10}$
 & ($5.20^{+0.01+2.94+0.29}_{-0.01-1.44-0.28}){\times}10^{-13}$
 & ($7.57^{+0.01+4.27+0.42}_{-0.01-2.10-0.40}){\times}10^{-12}$
   \end{tabular}
   \end{ruledtabular}
   \end{sidewaystable}

  \begin{table}[h]
  \caption{Classification of the nonleptonic charmonium weak decays.}
   \label{tab:case}
  \begin{ruledtabular}
  \begin{tabular}{lll}
  case & parametere & CKM factor \\ \hline
  1-a  & $a_{1}$ & ${\vert}V_{ud}V_{cs}^{\ast}{\vert}$ ${\sim}$ $1$ \\
  1-b  & $a_{1}$ & ${\vert}V_{ud}V_{cd}^{\ast}{\vert}$,
                   ${\vert}V_{us}V_{cs}^{\ast}{\vert}$ ${\sim}$ ${\lambda}$ \\
  1-c  & $a_{1}$ & ${\vert}V_{us}V_{cd}^{\ast}{\vert}$ ${\sim}$ ${\lambda}^{2}$ \\ \hline
  2-a  & $a_{2}$ & ${\vert}V_{ud}V_{cs}^{\ast}{\vert}$ ${\sim}$ $1$ \\
  2-b  & $a_{2}$ & ${\vert}V_{ud}V_{cd}^{\ast}{\vert}$,
                   ${\vert}V_{us}V_{cs}^{\ast}{\vert}$ ${\sim}$ ${\lambda}$ \\
  2-c  & $a_{2}$ & ${\vert}V_{us}V_{cd}^{\ast}{\vert}$ ${\sim}$ ${\lambda}^{2}$
  \end{tabular}
  \end{ruledtabular}
  \end{table}

  The input parameters including the CKM Wolfenstein parameters,
  decay constants, Gegenbauer moments and so on, are
  collected in Table \ref{tab:in}.
  If not specified explicitly, we will take their central values as
  the default inputs.
  Our numerical results on branching ratios for the nonleptonic
  two-body ${\psi}(1S,2S)$, ${\eta}_{c}(1S,2S)$
  ${\to}$ $DM$ weak decays are displayed in Tables \ref{tab:psi}
  and \ref{tab:etac},
  where the uncertainties of this work come from the CKM parameters,
  the renormalization scale ${\mu}$ $=$ $(1{\pm}0.2)m_{c}$,
  hadronic parameters including decay constants and Gegenbauer moments,
  respectively.
  For comparison, previous results on the $J/{\psi}$ weak decays
  \cite{epjc55,ijma14,adv2013} with parameters
  $a_{1}$ $=$ $1.26$ and $a_{2}$ $=$ $-0.51$
  are also listed in Table \ref{tab:psi}.
  The following are some comments.

  (1)
  There are some differences among the estimates of
  branching ratios for the $J/{\psi}$ ${\to}$ $DM$ weak
  decays (see the numbers in Table \ref{tab:psi}).
  These inconsistencies among previous works,
  although the same values of parameters $a_{1,2}$ are used,
  come principally from different values of form factors.
  Our results are generally in line with the numbers in
  columns ``A'' and ``B'' which are favored by Ref. \cite{adv2013}.

  (2)
  Branching ratios for the $J/{\psi}$ weak decay are about two
  or more times as large as those for ${\psi}(2S)$ decay into
  the same final states, because the decay width of ${\psi}(2S)$
  is about three times as large as that of $J/{\psi}$.

  (3)
  Due to the relatively small decay width and relatively large
  space phases for the ${\eta}_{c}(2S)$ decay, branching ratios
  for the ${\eta}_{c}(2S)$ weak decay are some five (ten) or more
  times as large as those for the ${\eta}_{c}(1S)$ weak decay
  into the same $DP$ ($DV$) final states.

  (4)
  Among the ${\psi}(1S,2S)$ and ${\eta}_{c}(1S,2S)$ mesons,
  the ${\eta}_{c}(1S)$ has a maximal decay width
  and a minimal mass resulting in a small phase space,
  while $J/{\psi}$ has a minimal decay width.
  These facts lead to the smallest [or the largest] branching
  ratio for ${\eta}_{c}(1S)$ [or $J/{\psi}$] weak decay
  among ${\psi}(1S,2S)$, ${\eta}_{c}(1S,2S)$
  weak decays into the same final states.

  (5)
  Compared with the ${\psi}(1S,2S)$ ${\to}$ $DV$ decays,
  the corresponding ${\psi}(1S,2S)$ ${\to}$ $DP$ decays,
  where $P$ and $V$ have the same flavor structures,
  are suppressed by the orbital angular momentum,
  so have relatively small branching ratios.
  There are some approximative
  relations ${\cal B}r(J/{\psi}{\to}DV)$
  ${\approx}$ $3{\cal B}r(J/{\psi}{\to}DP)$ and
  ${\cal B}r({\psi}(2S){\to}DV)$
  ${\approx}$ $3{\cal B}r({\psi}(2S){\to}DP)$.

  (6)
  According to the CKM factors and parameters $a_{1,2}$,
  nonleptonic charmonium weak decays could be subdivided
  into six cases (see Table \ref{tab:case}).
  Case ``i-a'' is the Cabibbo-favored one, so it generally has
  large branching ratios relative to case ``i-b'' and ``i-c''.
  The $a_{2}$-dominated charmonium weak decays are
  suppressed by a color factor relative to the
  $a_{1}$-dominated ones.
  Hence, the charmonium weak decays into the $D_{s}{\rho}$ and
  $D_{s}{\pi}$ final states belonging to case ``1-a'' usually
  have relatively large branching ratios;
  the charmonium weak decays into the $\overline{D}_{u}^{0}K^{{\ast}0}$
  final states belonging to case ``2-c'' usually have relatively
  small branching ratios.
  In addition, the branching ratio of case ``2-a'' (or ``2-b'')
  is usually larger than that of case ``1-b'' (or ``1-c'')
  due to ${\vert}a_{2}/a_{1}{\vert}$ ${\ge}$ ${\lambda}$.

  (7)
  Branching ratios for the Cabibbo favored ${\psi}(1S,2S)$ ${\to}$
  $D_{s}^{-}{\rho}^{+}$, $D_{s}^{-}{\pi}^{+}$,
  $\overline{D}_{u}^{0}\overline{K}^{{\ast}0}$ decays
  can reach up to $10^{-10}$, which might be measurable
  in the forthcoming days.
  For example, the $J/{\psi}$ production cross section
  can reach up to a few ${\mu}b$ with the LHCb and ALICE
  detectors at LHC \cite{1509.00771,npb850.387}.
  Therefore, over $10^{12}$ $J/{\psi}$ samples are in principle
  available per 100 ${\rm fb}^{-1}$ data collected by LHCb
  and ALICE, corresponding to a few tens of $J/{\psi}$ ${\to}$
  $D_{s}^{-}{\rho}^{+}$, $D_{s}^{-}{\pi}^{+}$,
  $\overline{D}_{u}^{0}\overline{K}^{{\ast}0}$
  events for about 10\% reconstruction efficiency.

  (8)
  There is a large cancellation between the CKM factor
  $V_{ud}V_{cd}^{\ast}$ and $V_{us}V_{cs}^{\ast}$, which
  results in a very small branching ratio for charmonium
  weak decays into $D_{u}{\eta}^{\prime}$ state.

  (9)
  There are many uncertainties on our results.
  The first uncertainty from the CKM factors is small due to
  high precision on the Wolfenstein parameter ${\lambda}$
  with only 0.3\% relative errors now \cite{pdg}.
  The second uncertainty from the renormalization scale ${\mu}$
  could, in principle, be reduced by the inclusion of higher
  order ${\alpha}_{s}$ corrections. 
  For example, it has been showed \cite{nnlo} that tree amplitudes
  incorporating with the NNLO corrections are relatively
  less sensitive to the renormalization scale than
  the NLO amplitudes.
  The third uncertainty comes from hadronic parameters,
  which is expected to be cancelled or reduced
  with the relative ratio of branching ratios.

  (10)
  The numbers in Table \ref{tab:psi} and \ref{tab:etac}
  just provide an order of magnitude estimate.
  Many other factors, such as the final state interactions,
  $q^{2}$ dependence of form factors and so on,
  which are not considered here,
  deserve many dedicated studies.

  \section{Summary}
  \label{sec04}
  With the anticipation of abundant data samples on charmonium
  at high-luminosity heavy-flavor experiments,
  we studied the nonleptonic two-body
  ${\psi}(1S,2S)$ and ${\eta}_{c}(1S,2S)$ weak decays
  into one ground-state charmed meson plus one ground-state light
  meson based on the low energy effective Hamiltonian.
  By considering QCD radiative corrections to hadronic
  matrix elements of tree operators, we got the effective
  coefficients $a_{1,2}$ containing partial information of strong phases.
  The magnitude of $a_{1,2}$ agrees comfortably with those
  used in previous works \cite{epjc55,plb252,ijma14,adv2013}.
  The transition form factors between the charmonium and
  charmed meson are calculated by using the nonrelativistic wave
  functions with isotropic harmonic oscillator potential.
  Branching ratios for ${\psi}(1S,2S)$, ${\eta}_{c}(1S,2S)$
  ${\to}$ $DM$ decays are estimated roughly.
  It is found that the Cabibbo favored ${\psi}(1S,2S)$ ${\to}$
  $D_{s}^{-}{\rho}^{+}$, $D_{s}^{-}{\pi}^{+}$,
  $\overline{D}_{u}^{0}\overline{K}^{{\ast}0}$ decays have
  large branching ratios ${\gtrsim}$ $10^{-10}$,
  which are promisingly detected in the forthcoming years.

  \section*{Acknowledgments}
  The work is supported by the National Natural Science Foundation
  of China (Grant Nos. 11547014, 11275057, 11475055, U1232101 and U1332103).
  We thank the referees for their helpful comments.

  \begin{appendix}
  \section{The amplitudes for ${\psi}$ ${\to}$ $DM$ decays}
  \label{app01}
  \begin{eqnarray}
  {\cal A}({\psi}{\to}D_{s}^{-}{\pi}^{+})
  &=&
   \sqrt{2}\, G_{F}\, m_{\psi}\, ({\epsilon}_{\psi}{\cdot}p_{\pi})\,
   f_{\pi}\, A_{0}^{{\psi}{\to}D_{s}}\,
   V_{cs}^{\ast}\, V_{ud}\, a_{1}
   \label{amp-psi-cs-pi}, \\
  {\cal A}({\psi}{\to}D_{s}^{-}K^{+})
  &=&
   \sqrt{2}\, G_{F}\, m_{\psi}\, ({\epsilon}_{\psi}{\cdot}p_{K})\,
   f_{K}\, A_{0}^{{\psi}{\to}D_{s}}\,
   V_{cs}^{\ast}\, V_{us}\, a_{1}
   \label{amp-psi-cs-k}, \\
  {\cal A}({\psi}{\to}D_{d}^{-}{\pi}^{+})
  &=&
   \sqrt{2}\, G_{F}\, m_{\psi}\, ({\epsilon}_{\psi}{\cdot}p_{\pi})\,
   f_{\pi}\, A_{0}^{{\psi}{\to}D_{d}}\,
   V_{cd}^{\ast}\, V_{ud}\, a_{1}
   \label{amp-psi-cd-pi}, \\
  {\cal A}({\psi}{\to}D_{d}^{-}K^{+})
  &=&
   \sqrt{2}\, G_{F}\, m_{\psi}\, ({\epsilon}_{\psi}{\cdot}p_{K})\,
   f_{K}\, A_{0}^{{\psi}{\to}D_{d}}\,
   V_{cd}^{\ast}\, V_{us}\, a_{1}
   \label{amp-psi-cd-k}, \\
  {\cal A}({\psi}{\to}\overline{D}_{u}^{0}{\pi}^{0})
  &=&
  -G_{F}\, m_{\psi}\, ({\epsilon}_{\psi}{\cdot}p_{\pi})\,
   f_{\pi}\, A_{0}^{{\psi}{\to}D_{u}}\,
   V_{cd}^{\ast}\, V_{ud}\, a_{2}
   \label{amp-psi-cu-pi}, \\
  {\cal A}({\psi}{\to}\overline{D}_{u}^{0}K^{0})
  &=&
   \sqrt{2}\, G_{F}\, m_{\psi}\, ({\epsilon}_{\psi}{\cdot}p_{K})\,
   f_{K}\, A_{0}^{{\psi}{\to}D_{u}}\,
   V_{cd}^{\ast}\, V_{us}\, a_{2}
   \label{amp-psi-cu-kz}, \\
  {\cal A}({\psi}{\to}\overline{D}_{u}^{0}\overline{K}^{0})
  &=&
   \sqrt{2}\, G_{F}\, m_{\psi}\, ({\epsilon}_{\psi}{\cdot}p_{K})\,
   f_{K}\, A_{0}^{{\psi}{\to}D_{u}}\,
   V_{cs}^{\ast}\, V_{ud}\, a_{2}
   \label{amp-psi-cu-kzb}, \\
  {\cal A}({\psi}{\to}\overline{D}_{u}^{0}{\eta}_{q})
  &=&
   G_{F}\, m_{\psi}\, ({\epsilon}_{\psi}{\cdot}p_{{\eta}_{q}})\,
   f_{{\eta}_{q}}\, A_{0}^{{\psi}{\to}D_{u}}\,
   V_{cd}^{\ast}\, V_{ud}\, a_{2}
   \label{amp-psi-cu-etaq}, \\
  {\cal A}({\psi}{\to}\overline{D}_{u}^{0}{\eta}_{s})
  &=&
   \sqrt{2}\, G_{F}\, m_{\psi}\, ({\epsilon}_{\psi}{\cdot}p_{{\eta}_{s}})\,
   f_{{\eta}_{s}}\, A_{0}^{{\psi}{\to}D_{u}}\,
   V_{cs}^{\ast}\, V_{us}\, a_{2}
   \label{amp-psi-cu-etas}, \\
  {\cal A}({\psi}{\to}\overline{D}_{u}^{0}{\eta})
  &=&
  {\cos}{\phi}\,{\cal A}({\psi}{\to}\overline{D}_{u}^{0}{\eta}_{q})
 -{\sin}{\phi}\,{\cal A}({\psi}{\to}\overline{D}_{u}^{0}{\eta}_{s})
   \label{amp-psi-cu-eta}, \\
  {\cal A}({\psi}{\to}\overline{D}_{u}^{0}{\eta}^{\prime})
  &=&
  {\sin}{\phi}\,{\cal A}({\psi}{\to}\overline{D}_{u}^{0}{\eta}_{q})
 +{\cos}{\phi}\,{\cal A}({\psi}{\to}\overline{D}_{u}^{0}{\eta}_{s})
   \label{amp-psi-cu-etap}.
  \end{eqnarray}
  \begin{eqnarray}
  {\cal A}({\psi}{\to}D_{s}^{-}{\rho}^{+})
  &=&
   -i\,\frac{G_{F}}{\sqrt{2}}\, f_{\rho}\, m_{\rho}\,
   V_{cs}^{\ast}\, V_{ud}\, a_{1}\, \Big\{
   ({\epsilon}_{\rho}^{\ast}{\cdot}{\epsilon}_{\psi})\,
   (m_{\psi}+m_{D_{s}})\, A_{1}^{{\psi}{\to}D_{s}}
   \nonumber \\ & & \hspace{-15mm}
   + ({\epsilon}_{\rho}^{\ast}{\cdot}p_{\psi})\,
  ({\epsilon}_{\psi}{\cdot}p_{\rho})\,
   \frac{2\, A_{2}^{{\psi}{\to}D_{s}}}{ m_{\psi}+m_{D_{s}} }
  -i\,{\epsilon}_{{\mu}{\nu}{\alpha}{\beta}}\,
  {\epsilon}_{\rho}^{{\ast}{\mu}}\,{\epsilon}_{\psi}^{\nu}\,
  p_{\rho}^{\alpha}\,p_{\psi}^{\beta}\,
  \frac{2\, V^{{\psi}{\to}D_{s}}}{ m_{\psi}+m_{D_{s}} }
   \Big\}
   \label{amp-psi-cs-rho}, \\
  {\cal A}({\psi}{\to}D_{s}^{-}K^{{\ast}+})
  &=&
   -i\,\frac{G_{F}}{\sqrt{2}}\, f_{K^{\ast}}\, m_{K^{\ast}}\,
   V_{cs}^{\ast}\, V_{us}\, a_{1}\, \Big\{
   ({\epsilon}_{K^{\ast}}^{\ast}{\cdot}{\epsilon}_{\psi})\,
   (m_{\psi}+m_{D_{s}})\, A_{1}^{{\psi}{\to}D_{s}}
   \nonumber \\ & & \hspace{-15mm}
   + ({\epsilon}_{K^{\ast}}^{\ast}{\cdot}p_{\psi})\,
  ({\epsilon}_{\psi}{\cdot}p_{K^{\ast}})\,
   \frac{2\, A_{2}^{{\psi}{\to}D_{s}}}{ m_{\psi}+m_{D_{s}} }
  -i\,{\epsilon}_{{\mu}{\nu}{\alpha}{\beta}}\,
  {\epsilon}_{K^{\ast}}^{{\ast}{\mu}}\,{\epsilon}_{\psi}^{\nu}\,
  p_{K^{\ast}}^{\alpha}\,p_{\psi}^{\beta}\,
  \frac{2\, V^{{\psi}{\to}D_{s}}}{ m_{\psi}+m_{D_{s}} }
   \Big\}
   \label{amp-psi-cs-kv}, \\
  {\cal A}({\psi}{\to}D_{d}^{-}{\rho}^{+})
  &=&
  -i\,\frac{G_{F}}{\sqrt{2}}\, f_{\rho}\, m_{\rho}\,
   V_{cd}^{\ast}\, V_{ud}\, a_{1}\, \Big\{
   ({\epsilon}_{\rho}^{\ast}{\cdot}{\epsilon}_{\psi})\,
   (m_{\psi}+m_{D_{d}})\, A_{1}^{{\psi}{\to}D_{d}}
   \nonumber \\ & & \hspace{-15mm}
   + ({\epsilon}_{\rho}^{\ast}{\cdot}p_{\psi})\,
  ({\epsilon}_{\psi}{\cdot}p_{\rho})\,
   \frac{2\, A_{2}^{{\psi}{\to}D_{d}}}{ m_{\psi}+m_{D_{d}} }
  -i\,{\epsilon}_{{\mu}{\nu}{\alpha}{\beta}}\,
  {\epsilon}_{\rho}^{{\ast}{\mu}}\,{\epsilon}_{\psi}^{\nu}\,
  p_{\rho}^{\alpha}\,p_{\psi}^{\beta}\,
  \frac{2\, V^{{\psi}{\to}D_{d}}}{ m_{\psi}+m_{D_{d}} }
   \Big\}
   \label{amp-psi-cd-rho}, \\
  {\cal A}({\psi}{\to}D_{d}^{-}K^{{\ast}+})
  &=&
   -i\,\frac{G_{F}}{\sqrt{2}}\, f_{K^{\ast}}\, m_{K^{\ast}}\,
   V_{cd}^{\ast}\, V_{us}\, a_{1}\, \Big\{
   ({\epsilon}_{K^{\ast}}^{\ast}{\cdot}{\epsilon}_{\psi})\,
   (m_{\psi}+m_{D_{d}})\, A_{1}^{{\psi}{\to}D_{d}}
   \nonumber \\ & & \hspace{-15mm}
   + ({\epsilon}_{K^{\ast}}^{\ast}{\cdot}p_{\psi})\,
  ({\epsilon}_{\psi}{\cdot}p_{K^{\ast}})\,
   \frac{2\, A_{2}^{{\psi}{\to}D_{d}}}{ m_{\psi}+m_{D_{d}} }
  -i\,{\epsilon}_{{\mu}{\nu}{\alpha}{\beta}}\,
  {\epsilon}_{K^{\ast}}^{{\ast}{\mu}}\,{\epsilon}_{\psi}^{\nu}\,
  p_{K^{\ast}}^{\alpha}\,p_{\psi}^{\beta}\,
  \frac{2\, V^{{\psi}{\to}D_{d}}}{ m_{\psi}+m_{D_{d}} }
   \Big\}
   \label{amp-psi-cd-kv}, \\
  {\cal A}({\psi}{\to}\overline{D}_{u}^{0}{\rho}^{0})
  &=&
   +i\,\frac{G_{F}}{2}\, f_{\rho}\, m_{\rho}\,
   V_{cd}^{\ast}\, V_{ud}\, a_{2}\, \Big\{
   ({\epsilon}_{\rho}^{\ast}{\cdot}{\epsilon}_{\psi})\,
   (m_{\psi}+m_{D_{u}})\, A_{1}^{{\psi}{\to}D_{u}}
   \nonumber \\ & & \hspace{-15mm}
   + ({\epsilon}_{\rho}^{\ast}{\cdot}p_{\psi})\,
  ({\epsilon}_{\psi}{\cdot}p_{\rho})\,
   \frac{2\, A_{2}^{{\psi}{\to}D_{u}}}{ m_{\psi}+m_{D_{u}} }
  -i\,{\epsilon}_{{\mu}{\nu}{\alpha}{\beta}}\,
  {\epsilon}_{\rho}^{{\ast}{\mu}}\,{\epsilon}_{\psi}^{\nu}\,
  p_{\rho}^{\alpha}\,p_{\psi}^{\beta}\,
   \frac{2\, V^{{\psi}{\to}D_{u}}}{ m_{\psi}+m_{D_{u}} }
   \Big\}
   \label{amp-psi-cu-rho}, \\
  {\cal A}({\psi}{\to}\overline{D}_{u}^{0}{\omega})
  &=&
   -i\,\frac{G_{F}}{2}\, f_{\omega}\, m_{\omega}\,
   V_{cd}^{\ast}\, V_{ud}\, a_{2}\, \Big\{
   ({\epsilon}_{\omega}^{\ast}{\cdot}{\epsilon}_{\psi})\,
   (m_{\psi}+m_{D_{u}})\, A_{1}^{{\psi}{\to}D_{u}}
   \nonumber \\ & & \hspace{-15mm}
   + ({\epsilon}_{\omega}^{\ast}{\cdot}p_{\psi})\,
  ({\epsilon}_{\psi}{\cdot}p_{\omega})\,
   \frac{2\, A_{2}^{{\psi}{\to}D_{u}}}{ m_{\psi}+m_{D_{u}} }
  -i\,{\epsilon}_{{\mu}{\nu}{\alpha}{\beta}}\,
  {\epsilon}_{\omega}^{{\ast}{\mu}}\,{\epsilon}_{\psi}^{\nu}\,
  p_{\omega}^{\alpha}\,p_{\psi}^{\beta}\,
   \frac{2\, V^{{\psi}{\to}D_{u}}}{ m_{\psi}+m_{D_{u}} }
   \Big\}
   \label{amp-psi-cu-omega}, \\
  {\cal A}({\psi}{\to}\overline{D}_{u}^{0}{\phi})
  &=&
   -i\,\frac{G_{F}}{\sqrt{2}}\, f_{\phi}\, m_{\phi}\,
   V_{cs}^{\ast}\, V_{us}\, a_{2}\, \Big\{
   ({\epsilon}_{\phi}^{\ast}{\cdot}{\epsilon}_{\psi})\,
   (m_{\psi}+m_{D_{u}})\, A_{1}^{{\psi}{\to}D_{u}}
   \nonumber \\ & & \hspace{-15mm}
   + ({\epsilon}_{\phi}^{\ast}{\cdot}p_{\psi})\,
  ({\epsilon}_{\psi}{\cdot}p_{\phi})\,
   \frac{2\, A_{2}^{{\psi}{\to}D_{u}}}{ m_{\psi}+m_{D_{u}} }
  -i\,{\epsilon}_{{\mu}{\nu}{\alpha}{\beta}}\,
  {\epsilon}_{\phi}^{{\ast}{\mu}}\,{\epsilon}_{\psi}^{\nu}\,
  p_{\phi}^{\alpha}\,p_{\psi}^{\beta}\,
   \frac{2\, V^{{\psi}{\to}D_{u}}}{ m_{\psi}+m_{D_{u}} }
   \Big\}
   \label{amp-psi-cu-phi}, \\
  {\cal A}({\psi}{\to}\overline{D}_{u}^{0}K^{{\ast}0})
  &=&
    -i\,\frac{G_{F}}{\sqrt{2}}\, f_{K^{\ast}}\, m_{K^{\ast}}\,
   V_{cd}^{\ast}\, V_{us}\, a_{2}\, \Big\{
   ({\epsilon}_{K^{\ast}}^{\ast}{\cdot}{\epsilon}_{\psi})\,
   (m_{\psi}+m_{D_{u}})\, A_{1}^{{\psi}{\to}D_{u}}
   \nonumber \\ & & \hspace{-15mm}
   + ({\epsilon}_{K^{\ast}}^{\ast}{\cdot}p_{\psi})\,
  ({\epsilon}_{\psi}{\cdot}p_{K^{\ast}})\,
   \frac{2\, A_{2}^{{\psi}{\to}D_{u}}}{ m_{\psi}+m_{D_{u}} }
  -i\,{\epsilon}_{{\mu}{\nu}{\alpha}{\beta}}\,
  {\epsilon}_{K^{\ast}}^{{\ast}{\mu}}\,{\epsilon}_{\psi}^{\nu}\,
  p_{K^{\ast}}^{\alpha}\,p_{\psi}^{\beta}\,
   \frac{2\, V^{{\psi}{\to}D_{u}}}{ m_{\psi}+m_{D_{u}} }
   \Big\}
   \label{amp-psi-cu-kvz}, \\
  {\cal A}({\psi}{\to}\overline{D}_{u}^{0}\overline{K}^{{\ast}0})
  &=&
  -i\,\frac{G_{F}}{\sqrt{2}}\, f_{K^{\ast}}\,
  m_{K^{\ast}}\, V_{cs}^{\ast}\, V_{ud}\, a_{2}\, \Big\{
   ({\epsilon}_{K^{\ast}}^{\ast}{\cdot}{\epsilon}_{\psi})\,
   (m_{\psi}+m_{D_{u}})\, A_{1}^{{\psi}{\to}D_{u}}
   \nonumber \\ & & \hspace{-15mm}
   + ({\epsilon}_{K^{\ast}}^{\ast}{\cdot}p_{\psi})\,
  ({\epsilon}_{\psi}{\cdot}p_{K^{\ast}})\,
   \frac{2\, A_{2}^{{\psi}{\to}D_{u}}}{ m_{\psi}+m_{D_{u}} }
  -i\,{\epsilon}_{{\mu}{\nu}{\alpha}{\beta}}\,
  {\epsilon}_{K^{\ast}}^{{\ast}{\mu}}\,{\epsilon}_{\psi}^{\nu}\,
  p_{K^{\ast}}^{\alpha}\,p_{\psi}^{\beta}\,
   \frac{2\, V^{{\psi}{\to}D_{u}}}{ m_{\psi}+m_{D_{u}} }
   \Big\}
   \label{amp-psi-cu-kvzb}.
  \end{eqnarray}

  \section{The amplitudes for the ${\eta}_{c}$ ${\to}$ $DM$ decays}
  \label{app02}
  \begin{eqnarray}
  {\cal A}({\eta}_{c}{\to}D_{s}^{-}{\pi}^{+})
  &=&
   i\frac{G_{F}}{\sqrt{2}}\,
   (m_{{\eta}_{c}}^{2}-m_{D_{s}}^{2})\,
   f_{\pi}\, F_{0}^{{\eta}_{c}{\to}D_{s}}\,
   V_{ud}\, V_{cs}^{\ast}\, a_{1}
   \label{amp-etac-cs-pi}, \\
  {\cal A}({\eta}_{c}{\to}D_{s}^{-}K^{+})
  &=&
   i\frac{G_{F}}{\sqrt{2}}\,
   (m_{{\eta}_{c}}^{2}-m_{D_{s}}^{2})\,
   f_{K}\, F_{0}^{{\eta}_{c}{\to}D_{s}}\,
   V_{us}\, V_{cs}^{\ast}\, a_{1}
   \label{amp-etac-cs-k}, \\
  {\cal A}({\eta}_{c}{\to}D_{d}^{-}{\pi}^{+})
  &=&
   i\frac{G_{F}}{\sqrt{2}}\,
   (m_{{\eta}_{c}}^{2}-m_{D_{d}}^{2})\,
    f_{\pi}\, F_{0}^{{\eta}_{c}{\to}D_{d}}\,
   V_{ud}\, V_{cd}^{\ast}\, a_{1}
   \label{amp-etac-cd-pi}, \\
  {\cal A}({\eta}_{c}{\to}D_{d}^{-}K^{+})
  &=&
   i\frac{G_{F}}{\sqrt{2}}\,
   (m_{{\eta}_{c}}^{2}-m_{D_{d}}^{2})\,
   f_{K}\, F_{0}^{{\eta}_{c}{\to}D_{d}}\,
   V_{us}\, V_{cd}^{\ast}\, a_{1}
   \label{amp-etac-cd-k}, \\
  {\cal A}({\eta}_{c}{\to}\overline{D}_{u}^{0}{\pi}^{0})
  &=&
  -i\frac{G_{F}}{2}\,
   (m_{{\eta}_{c}}^{2}-m_{D_{u}}^{2})\,
   f_{\pi}\, F_{0}^{{\eta}_{c}{\to}D_{u}}\,
   V_{ud}\, V_{cd}^{\ast}\, a_{2}
   \label{amp-etac-cu-pi}, \\
  {\cal A}({\eta}_{c}{\to}\overline{D}_{u}^{0}K^{0})
  &=&
   i\frac{G_{F}}{\sqrt{2}}\,
   (m_{{\eta}_{c}}^{2}-m_{D_{u}}^{2})\,
   f_{K}\, F_{0}^{{\eta}_{c}{\to}D_{u}}\,
   V_{us}\, V_{cd}^{\ast}\, a_{2}
   \label{amp-etac-cu-kz}, \\
  {\cal A}({\eta}_{c}{\to}\overline{D}_{u}^{0}\overline{K}^{0})
  &=&
   i\frac{G_{F}}{\sqrt{2}}\,
   (m_{{\eta}_{c}}^{2}-m_{D_{u}}^{2})\,
   f_{K}\, F_{0}^{{\eta}_{c}{\to}D_{u}}\,
   V_{ud}\, V_{cs}^{\ast}\, a_{2}
   \label{amp-etac-cu-kzb}, \\
  {\cal A}({\eta}_{c}{\to}\overline{D}_{u}^{0}{\eta}_{q})
  &=&
  i\frac{G_{F}}{2}\,
   (m_{{\eta}_{c}}^{2}-m_{D_{u}}^{2})\,
   f_{{\eta}_{q}}\, F_{0}^{{\eta}_{c}{\to}D_{u}}\,
   V_{ud}\, V_{cd}^{\ast}\, a_{2}
   \label{amp-etac-cu-etaq}, \\
  {\cal A}({\eta}_{c}{\to}\overline{D}_{u}^{0}{\eta}_{s})
  &=&
  i\frac{G_{F}}{\sqrt{2}}\,
   (m_{{\eta}_{c}}^{2}-m_{D_{u}}^{2})\,
   f_{{\eta}_{s}}\, F_{0}^{{\eta}_{c}{\to}D_{u}}\,
   V_{us}\, V_{cs}^{\ast}\, a_{2}
   \label{amp-etac-cu-etas}, \\
  {\cal A}({\eta}_{c}{\to}\overline{D}_{u}^{0}{\eta})
  &=&
  {\cos}{\phi}\,{\cal A}({\eta}_{c}{\to}\overline{D}_{u}^{0}{\eta}_{q})
 -{\sin}{\phi}\,{\cal A}({\eta}_{c}{\to}\overline{D}_{u}^{0}{\eta}_{s})
   \label{amp-etac-cu-eta}, \\
  {\cal A}({\eta}_{c}{\to}\overline{D}_{u}^{0}{\eta}^{\prime})
  &=&
  {\sin}{\phi}\,{\cal A}({\eta}_{c}{\to}\overline{D}_{u}^{0}{\eta}_{q})
 +{\cos}{\phi}\,{\cal A}({\eta}_{c}{\to}\overline{D}_{u}^{0}{\eta}_{s})
   \label{amp-etac-cu-etap}, \\
  {\cal A}({\eta}_{c}{\to}D_{s}^{-}{\rho}^{+})
  &=&
   \sqrt{2}\, G_{F}\, m_{\rho}\,
  ({\epsilon}_{\rho}^{\ast}{\cdot}p_{{\eta}_{c}})\,
  f_{\rho}\, F_{1}^{{\eta}_{c}{\to}D_{s}}\,
   V_{ud}\, V_{cs}^{\ast}\, a_{1}
   \label{amp-etac-cs-rho}, \\
  {\cal A}({\eta}_{c}{\to}D_{s}^{-}K^{{\ast}+})
  &=&
   \sqrt{2}\, G_{F}\, m_{K^{\ast}}\,
  ({\epsilon}_{K^{\ast}}^{\ast}{\cdot}p_{{\eta}_{c}})\,
   f_{K^{\ast}}\, F_{1}^{{\eta}_{c}{\to}D_{s}}\,
   V_{us}\, V_{cs}^{\ast}\, a_{1}
   \label{amp-etac-cs-kv}, \\
  {\cal A}({\eta}_{c}{\to}D_{d}^{-}{\rho}^{+})
  &=&
  \sqrt{2}\, G_{F}\, m_{\rho}\,
  ({\epsilon}_{\rho}^{\ast}{\cdot}p_{{\eta}_{c}})\,
   f_{\rho}\, F_{1}^{{\eta}_{c}{\to}D_{d}}\,
   V_{ud}\, V_{cd}^{\ast}\, a_{1}
   \label{amp-etac-cd-rho}, \\
  {\cal A}({\eta}_{c}{\to}D_{d}^{-}K^{{\ast}+})
  &=&
   \sqrt{2}\, G_{F}\, m_{K^{\ast}}\,
  ({\epsilon}_{K^{\ast}}^{\ast}{\cdot}p_{{\eta}_{c}})\,
   f_{K^{\ast}}\, F_{1}^{{\eta}_{c}{\to}D_{d}}\,
   V_{us}\, V_{cd}^{\ast}\, a_{1}
   \label{amp-etac-cd-kv}, \\
  {\cal A}({\eta}_{c}{\to}\overline{D}_{u}^{0}{\rho}^{0})
  &=&
   - G_{F}\, m_{\rho}\,
  ({\epsilon}_{\rho}^{\ast}{\cdot}p_{{\eta}_{c}})\,
   f_{\rho}\, F_{1}^{{\eta}_{c}{\to}D_{u}}\,
   V_{ud}\, V_{cd}^{\ast}\, a_{2}
   \label{amp-etac-cu-rho}, \\
  {\cal A}({\eta}_{c}{\to}\overline{D}_{u}^{0}{\omega})
  &=&
   G_{F}\, m_{\omega}\,
  ({\epsilon}_{\omega}^{\ast}{\cdot}p_{{\eta}_{c}})\,
   f_{\omega}\, F_{1}^{{\eta}_{c}{\to}D_{u}}\,
   V_{ud}\, V_{cd}^{\ast}\, a_{2}
   \label{amp-etac-cu-omega}, \\
  {\cal A}({\eta}_{c}{\to}\overline{D}_{u}^{0}{\phi})
  &=&
  \sqrt{2}\, G_{F}\, m_{\phi}\,
  ({\epsilon}_{\phi}^{\ast}{\cdot}p_{{\eta}_{c}})\,
   f_{\phi}\, F_{1}^{{\eta}_{c}{\to}D_{u}}\,
   V_{us}\, V_{cs}^{\ast}\, a_{2}
   \label{amp-etac-cu-phi}, \\
  {\cal A}({\eta}_{c}{\to}\overline{D}_{u}^{0}K^{{\ast}0})
  &=&
   \sqrt{2}\, G_{F}\, m_{K^{\ast}}\,
  ({\epsilon}_{K^{\ast}}^{\ast}{\cdot}p_{{\eta}_{c}})\,
   f_{K^{\ast}}\, F_{1}^{{\eta}_{c}{\to}D_{u}}\,
   V_{us}\, V_{cd}^{\ast}\, a_{2}
   \label{amp-etac-cu-kvz}, \\
  {\cal A}({\eta}_{c}{\to}\overline{D}_{u}^{0}\overline{K}^{{\ast}0})
  &=&
  \sqrt{2}\, G_{F}\, m_{K^{\ast}}\,
  ({\epsilon}_{K^{\ast}}^{\ast}{\cdot}p_{{\eta}_{c}})\,
   f_{K^{\ast}}\, F_{1}^{{\eta}_{c}{\to}D_{u}}\,
   V_{ud}\, V_{cs}^{\ast}\, a_{2}
   \label{amp-etac-cu-kvzb}.
  \end{eqnarray}
  \end{appendix}


  \end{document}